# The International Monetary Fund's intervention in education systems and its impact on children's chances of completing school

Adel Daoud[1,2, *],

(Version 22 December 2021)

w

1. Institute for Analytical Sociology, Department of Management and Engineering, Linköping University, Norrköping, Sweden.w
2. The Division of Data Science and Artificial Intelligence of the Department of Computer Science and Engineering, Chalmers University of Technology, Gothenburg, Sweden

* Corresponding author: adel.daoud@liu.se

Enabling children to acquire an education is one of the most effective means to reduce inequality, poverty, and ill-health globally. While in normal times a government controls its educational policies, during times of macroeconomic instability, that control may shift to supporting international organizations, such as the the International Monetary Fund (IMF). While much research has focused on which sectors has been affected by IMF policies, scholars have devoted little attention to the policy content of IMF interventions affecting the education sector and children's education outcomes: denoted IMF-education policies. This article evaluates the extent which IMF-education policies exist in all programs and how these policies and IMF programs affect children's likelihood of completing schools. While IMF-education policies have a small adverse effect yet statistically insignificant on children's probability of completing school, these policies moderate effect heterogeneity for IMF programs. The effect of IMF programs (joint set of policies) adversely effect children's chances of completing school by six percentage points. By analyzing how IMF-education policies but also how IMF programs affect the education sector in low- and middle-income countries, scholars will gain a deeper understanding of how such policies will likely affect downstream outcomes.

# Table of Contents

# 1    Introduction

Enabling children to acquire an education is one of the most effective means to reduce inequality, poverty, and ill-health globally (Banerjee and Duflo 2012; Deaton 2015; Sen 1999). Having an education is also a critical ingredient for empowering girls in their future endeavors and decreasing the gap of unequal life opportunities (Breen and Jonsson 2005). The likelihood of a child completing schools depends on not only micro factors such as family living-conditions but also macro factors: government's effectiveness, structure of the education system (e.g., private or public schools) (Conklin et al. 2018; Halleröd et al. 2013; Nandy, Daoud, and Gordon 2016; Ponce et al. 2017; Rothstein 2011), and the amount of social spending on its education spending. The quality of the education system determines to a considerable extent a child's probability of completing school (Abdullahi 2015). While in normal times a government controls its educational policies, during times of macroeconomic instability, that control may shift to supporting international organizations—eroding their sovereignty over domestic affairs. That shift in control is likely to lead to unexpected consequences, and in the worst-case jeopardizing children's chances to acquire an education (Alexander 2001; Buchmann 1996; Daoud et al. 2017; Daoud and Johansson 2020; Moosa and Moosa 2019; Nielsen 2006).

One of the most powerful international organizations is the International Monetary Fund (IMF). As one of the several organizations sanctioned by the United Nation, the IMF's role is to support governments in macroeconomics turmoil (Vreeland 2007). It provides this support in an *IMF program* that contains financial assistance but also a set of policy conditions. In exchange for its support, the IMF requires that the government to implement a set of policy conditions that the IMF considers addressing the route of these macroeconomics imbalances. These policies are generally of neoliberal nature: liberalizing trade, privatizing state-owned companies, and deregulating markets (Babb 2005). While much research has focused on which sectors has been affected by IMF policies (Daoud, Herlitz, and Subramanian 2020; Dreher 2005, 2006; Vreeland 2007), especially health spending, scholars have devoted little attention to the policy content of IMF interventions affecting the education sector and children's education outcomes. Existing studies find that IMF policies tend to decrease education spending (Stubbs et al. 2018), but it remains unclear the extent which the IMF explicitly targets education sector directly through what we call IMF-education policies, and how such targeting affects children's education outcomes. By *IMF-education policies* we mean a policy that explicitly and directly stipulates a change in the education system. An IMF-education policy exists predominately as one of several other IMF policies, bundled together in IMF programs. While several datasets exist on IMF programs (e.g. Vreeland 2007), only the IMF's Monitoring of Fund Arrangements database (MONA) offers disaggregated information about the content of these programs. Nonetheless, MONA has been shown to be incomplete (Arpac, Bird, and Mandilaras 2008; Daoud, Reinsberg, et al. 2019; IEO 2007; Kentikelenis, Stubbs, and King 2016). The IMF conditionality dataset improves on this incompleteness by capturing a range of policy areas. Nonetheless, none of these capture IMF-education policies.

By analyzing how IMF-education policies but also how IMF programs affect the education sector in low- and middle-income countries, scholars will gain a deeper understanding of how such policies will likely affect downstream outcomes (Stuckler and Basu 2013). These outcomes range from not only the number of children graduating but also the quality of university studies, number of young adults in employment, and similar outcomes. Policymakers will gain insights into the likely affects should their country implement IMF programs.

The purpose of our article is it to evaluate the extent which IMF-education policies exist in all programs and how these policies and IMF programs affect children's likelihood of completing schools. First, our analysis identifies which IMF policies contain explicit reference to the education sector across all IMF policies in the period 1985-2014. Besides evaluating extent, this identification shows the temporal and geographical spread of IMF-education policies in the past. This analysis reveals the content of IMF-education policies, and how many conditions contain explicit reference to education. Second, besides deepening the scholarly knowledge on how IMF-education policies affect countries around the world, our analysis supplies a dataset that quantifies such policies—supporting future research. Third, using this IMF-education data, our analysis evaluates the impact of IMF-education policies and IMF programs on children's chances of completing primary or secondary school. Our analysis relies on the Demographic and Health Survey (DHS) and Multiple Indicator Cluster Survey (MICS), measuring the living conditions for over one million children living across half-a-million families and 67 low and middle-income countries. These measurements are representative samples of about half the worlds population around the year 2000.

Our policy evaluation captures both the average treatment effect (ATE) and the conditional average treatment effect (CATE). Because IMF policies are likely to affect groups of children differently, merely reporting ATE mask that variation in effect—called *effect heterogeneity*. An additional CATE analysis unpacks this effect heterogeneity across children's country, family, and individual characteristics. To identify CATE, we use a data-driven approach relying on statistical models for causal inference and machine learning—algorithms that find patterns in the data (Athey and Imbens 2017; Daoud and Dubhashi 2021; Hill 2011; Kino et al. 2021; Mullainathan and Spiess 2017).

## 2 Method and data

### 2.1 The outcome: child-education deprivation

We rely on the Bristol method to define children's education outcomes, denotes as $Y$. The Bristol method is a deprivation approach developed by Peter Townsend, his team, and UNICEF in 2003. A child is defined as deprived of an education, if that child "had never been to school and were not currently attending school, in other words, no professional education of any kind." The outcome, child-education deprivation, is a binary variable where one means "deprived" and zero means "not deprived." This outcome applies to children 7–17 years old.

Our data consists of household data from the DHS and MICS. These surveys are cross-sectional and nationally representative household surveys conducted several in middle-income and low-income countries. As these two surveys have identical sampling frameworks, we combine their data (Corsi et al. 2012). Using interviews, the response rate for these surveys normally exceed 90%. National samples range from 4,000 to 30,000 households depending on the population size. The DHS (2011) provides further information about the survey design.

We selected surveys from countries that were sampled around the year 2000. The DHS rolls out these surveys at different time points for each country. Our pooled data spans 1995 to 2005. Our household sample captures 1,150,711 children, representative of about 2.8 billion households, or roughly 50 percent of the world's population in the year 2000. Figure 1 highlights the sample's geographical distribution of child poverty.

[Figure 1 about here]

## 2.2   The exposures: IMF-education policies and IMF programs

Our analysis covers two types of exposures denoted jointly as $W$: IMF-education policies $W_1$ and IMF programs $W_2$. As defined in the Introduction, *IMF programs* are a set of policies that IMF and government officials have signed and agreed to implement in the target country. These programs consist of a set of policies, ranging from a few to up to five dozen. These policies consist of a variety of interventions, from small (e.g., appointing an expert for a specific area) to large ones (e.g., privatizing state-owned companies).  IMF program is a binary variable. We collect these data from the IMF, as they are freely available.

By *IMF-education policy*, we mean a policy that explicitly targets a country's education system. Clear cases are when the IMF explicitly requires introducing education fees, privatizing schools, or targeting teachers' salaries (up or down). While an IMF policy may require that a government adjusts its fiscal spending to achieve balance, which is not an IMF-education policy as it does not directly reference the education system, but asking to reduce social or public spending is a borderline case. These are borderline cases, as public or social spending often consist or closely overlaps with education spending and priorities in the social system. Demanding a reduction in public or social spending pushes the agency from the IMF to the government in deciding whether the education system should be affected or not. Although these borderline cases are important determinants for changes in the education system, they are too general for our purposes. That general case is included in our definition of IMF programs. For our purposes, we select a more stringent definition of IMF-education policies. A policy condition must include at least one of the following terms to qualify as an IMF-education policy,

> "[Ee]duca|[Uu]niversit[y|ies]|[Ss]chool|[Pp]edagog|[Tt]eacher|[Pp]roffesor|[Ll]ectur|[Ss]tudent|[Pp]upil|[Cc]lassroom|[Cc]urricul|[Ll]earn|[Ac]adem"

We denote this list or dictionary as $D$. The symbol "[ ]" means the term can be either capitalized or not and "|" means or. Despite that the list is short, it is calibrated to effectively capture policy text predominately relevant for education systems. A broad term list risks capturing irrelevant policies. As described in the Method section, we then match these terms to the IMF conditionality dataset, covering all IMF programs between 1985 to 2014. These data are derived from 4,500 IMF-program documents and include 58,406 conditions across 131 countries.

For the 67-countries with household data, separate from the 131 countries, we collected additional policy-text data, to substantiate our statistical findings. With that collection, we analyzed IMF program in depth, as documented in the IMF's Executive Board Specials (EBS)—digitally and freely available at the IMF website. An EBS documents lay out the national context and the specific policy conditions that the IMF and the particular government have drafted for a final agreement. These are then submitted—often with no or only minor amendments—to the IMF executive board for formal approval. As an IMF program usually runs for three years, with an additional period of extension (Vreeland 2007), we searched EBS documents in a six-year window, set to three years before and after the starting date of an IMF program. This window ensures that we capture policy drafts and updates relevant for program implementation.

## 2.3 Method

### 2.3.1 Identifying IMF-education policies in policy text

We use natural language processing to identify IMF-education policies (Åkerström, Daoud, and Johansson 2019; Daoud, Reinsberg, et al. 2019). As the IMF conditionality data records the actual policy conditions as it is written in its EBS document, we capitalize on this to identify IMF-education policies. Call the policy-condition text or corpus for $C_i$, where $i$ is the number of policy conditions searched, that is, 58,406. Before the search, used standard cleaning procedures recommended in text mining (Jockers 2014), by removing numbers and special characters as those do not carry qualitative meaning.[1] We then processed $C_i$, resulting in a document-term (in our case, conditionality-word) matrix, where each is represented as a vector of words $w_{ik}$. The index $k$ captures the number of words in policy condition $i$.

We then use our dictionary $D$—the list of terms identifying IMF-education policy defined in the Data section—and apply it to a search function $f$. This function takes $D$ as input, matches with a regular-expression searches, each word in term by term, whether $w_{ik}$ matches with any of the terms in $D$. If it does match, then the output of $f$ is one and conclude that the policy is likely an IMF-education policy; otherwise, the output is zero. More formally, we write this as follows,

$$f(C_i) = \begin{cases} 1, & if \ w_{ik} \in D \\ 0, & if \ w_{ik} \notin D \end{cases}$$

After matching, we sampled the results to verify that the whole policy text does indeed directly affect the education system.

### 2.3.2 Statistical analysis: estimating ATE and CATE

We combine a policy evaluation and machine learning methodology to estimate the impact of IMF program on children's risk of education deprivation (Daoud and Dubhashi 2020). Machine learning constitutes a set of nonparametric algorithms. In contrast to other statistical techniques such as generalized linear models, these algorithms are designed to optimize prediction without specific programming rules (Hastie, Tibshirani, and Friedman 2009). However, when combined with a policy-evaluation techniques, these learning algorithms provide at least four capabilities that traditional estimation techniques lack. In the outline that follows, we use the term *learning algorithms* to refer specifically to those machine learning algorithms that have been adapted for policy evaluation (Athey and Imbens 2017; Athey, Tibshirani, and Wager 2019). At the core of these adapted algorithms—compared to traditional machine learning algorithms—lay the assumption of an experimental (ignorability by randomization) or quasi-experimental design (conditional ignorability). In our case, we draw on the quasi-experimental design that our conceptual framework provides.

**Expected individual counterfactuals.** Under an assumption of conditional ignorability, learning algorithms allow us to impute the expected counterfactuals for each child (Daoud and Dubhashi 2021; Künzel et al. 2018). Matching methods is another instance of nonparametric techniques, but they rely instead on finding comparable cases in-sample. In addition, matching methods use all the covariates at their disposal, even those with low statistical

---

[1] We tested to lemmatize and stemming the corpus, in contrast to keep the corpus as it is. After validation we decided to rely on regular expression for the text search, as that produce the most robust results (in the sense that it gave the most conservative and valid hits regarding food and agricultural issues)

relevance for the outcome, while learning algorithms use mainly those variables that are predictive of the outcome. As long as the treated and control populations are comparable (in line with the overlap assumption discussed above), imputing counterfactuals provides a flexible method to conduct policy evaluation. We capitalize on this and impute a counterfactual for each child in our sample. The difference between these counterfactuals is the individual-level treatment effect. The difference between these counterfactuals is the individual-level conditional average treatment effect, $\tau(x_i, f_i) = E[Y(1) - Y(0)|X = x_i, F = f_i]$. This effect is a function of the observed covariates, $X = x_i$, and the type of model selected, $F = f_i$.

**Inductive investigation of impact heterogeneity.** Learning algorithms find impact heterogeneity inductively. Traditional regression requires that the analysts specify interaction variables explicitly, guided by theory. To manage complexity, they tend to specify two-way or at most three-way interactions (Neumayer and Plümper 2017). For example, if a researcher hypothesized that the effect of IMF programs (*W*) differed by a child's age (*X*), they would specify an interaction: $Y = \beta_0 + \beta_1 W + \beta_2 X + \beta_3 W \cdot X + e$. They would then test this model against the data to verify whether $\beta_3$ is significantly different from zero. Learning algorithms, however, do not require such explicit programing. They estimate many different models—possibly thousands—using random cuts of the data, and thus suggest where impact heterogeneity is largest (Shiba et al. 2021). They produce such suggestions by providing a list of variables that were important in building these many different models. Child age, for instance, might be relevant in only one percent of these many models, whereas household wealth might have been used 50 percent of the times. Accordingly, our learning algorithm will provide such a list to identify the most important variables moderating IMF impact heterogeneity.

**Guarding against researcher bias.** Learning algorithms guard against the influence of outliers, p-hacking, and cherry-picking results. Although this capability does not eradicate researcher discretion, it renders such bias less influential. Traditional regression procedures tend to use all the data simultaneously to generate results. Machine learning procedures, however, split the data into training and testing sets. The algorithm is first applied to the training set to find optimal regularization parameters. Then the same model is used on the testing set to estimate the quantities of interest. This procedure finds an optimal balance in the bias-variance tradeoff (Hastie et al. 2009). We rely on this splitting procedure to produce our results.

**Flexible functional form.** Learning algorithms find an optimal and flexible functional form. A model with a flexible functional form—imagine adding more and more interaction terms to a simple OLS regression—will eventually overfit the data. That is, the model will predict the given sample perfectly, implying that it will predict new data poorly. In machine learning, a regularization expression limits how flexible a model can become, thus balancing the trade-off between bias in-sample and variation out-of-sample. Selecting optimal regularization parameters, called *empirical tuning,* is done using cross-validation. This validation procedure involves estimating a model on a portion of the sample, and then evaluating predictive performance in another sample. Such models have been used in a variety in applications in the social and behavioral sciences (Daoud, Kim, and Subramanian 2019; Kino et al. 2021). Relying on novel methodological developments in policy evaluation, we select a function class called *generalized random forest* (Athey et al. 2019; Nie and Wager 2018).

### 2.3.3 Generalized random forest

Generalized random forest (*grf*) adapts the family of random forest (RF) estimators (Breiman 2001) for efficient non-parametric estimation of causal effects (Athey et al. 2019). RF models learn ensembles of regression (or classification) trees, each fit a different resampled population and covariate set, to estimate and reduce model variance. Each tree learns a set of rules (e.g., Age > 5) which partition the population of units into different leaves of the tree. The predicted outcome for a new unit is the average of outcomes for observed units assigned to the same leaf; the prediction of the forest is the average of the predictions of all trees. A strength of non-parametric (or machine learning) estimators such as RF is that they are designed to optimize predictive accuracy on held-out data by trading off bias and variance through regularization, rather than learning the parameters of a fixed-size model (Hastie et al. 2009). For tree-based estimators, many heuristic regularization strategies exist, including limiting the depth or number of leaf nodes in each tree. Generally, growing more trees is preferable for out-of-sample generalization. In order to select such tuning parameters, we use sample splitting, evaluating the predictive accuracy on a randomly subsampled set of held-out data, never exposed to the model. *grf* uses a version of sample splitting called "out-of-bag predictions." As it randomly picks a subset of cases from the full sample—hence, the name *random* forest—to grow each tree, it does not use all cases for all trees. Out-of-bag prediction applies each specific case to only those trees in which the *grf* did not sample that case to grown those trees. This type of prediction is a more efficient way to use data.

When machine learning is applied naïvely to policy evaluation design, the bias-variance trade-off described above may induce *regularization bias* in treatment effects since minimization of predictive error is agnostic to the difference between a treatment variable and other covariates. For example, regularized generalized linear models tend to prefer smaller coefficients and may, therefore, bias treatment effects in either direction, depending on the correlation between treatment and covariates. To address this issue, researchers have developed so-called debiased machine learning estimators which decouple the estimation of treatment propensity and effect heterogeneity (Chernozhukov, Newey, and Robins 2018; Nie and Wager 2018). *grf* implements this structure through the R-learner framework, blending regression and propensity score methods to obtain debiased estimates of causal effects (Nie and Wager 2018). The R-learner builds on the treatment effect decomposition of Robinson (1988) which relates CATE and covariates through a regression of a transformed outcome based on the observed treatment assignment. To achieve statistical efficiency, the R-learner draws on the strength of residualization, in which regressors are fit to residuals, rather than outcomes themselves. In particular, the R-learner first fits one model of treatment propensity $E[T|X]$ and one of marginal effect $E[Y|X]$. To obtain de-biased effect estimates, the residuals of these models are then explained by a third model—a model of the CATE with respect to X. This approach is also called quasi-oracle estimation as it relies on fitting a model to imputations of the treatment effect for each unit.

Once an estimate of CATE—using all the covariates—has been obtained, the population may be clustered at different levels of granularity to compile average treatment effects within groups. *grf* uses cluster-robust errors that ensures that the standard errors are computed correctly and that they are less sensitive to outliers. In our study, the treatment—IMF programs—is assigned at the country-level. Clustered standard errors reflected this uncertainty when estimating the quantities of interest. Additionally, we may be interested in the ATE stratified only by educational level. In this case, we average CATE predictions over all individuals in each stratum, effectively coarsening our view of treatment effect heterogeneity. If the set of confounders includes more variables than education, this two-stage approach is necessary to deconfound treatment effect estimates. It has the added benefit that the CATE model does not need to be retrained if several different stratifications are of interest.

### 2.3.4 Handling selection bias

As the exposures of interest $W$ are not randomized, our analysis is an observational study. Any observational study runs the risk of having the results contaminated by confounding. A confounding $X$ is a common cause to both the exposure $W$ and outcome $Y$. If a statistical analysis does not adjust for $X$, the results will likely be biased, contaminated by spurious effects (Pearl 2012). Although selection into IMF programs (or individual policies) are not likely driven by children's education status, it may still be the case that governments select into these programs because its population lives in poverty. And poverty and lack of education are closely related. Thus, we still require adjusting for a set of country- and family-level confounders. Appendix A shows all sources and definitions.

The set of confounders are the following. **GDP growth**: societies with lower economic growth are more likely to become economically constrained, and ask for IMF credit. **Current account balance**: the higher the fiscal imbalance, the more likely the country is to ask for IMF help. **Log GDP per capita**: low-income countries tend to seek concessional IMF assistance, whereas middle-income countries with short-term economic disturbances following, for example, a currency crisis tend to ask for non-concessional loans. **Trade**: the sum of exports and imports of goods and services measured as a share of the gross domestic product. **Economic globalization index**: how integrated a nation is in the global economy will affect how many trade liberalization policies the IMF will aim to include in a program (Dreher 2006). **High inflation indicator**: a rapid price rise indicates structural imbalances in the economy. **Negative growth**: loss of GDP indicates economic downturn, not necessarily related to current account imbalances. Although we adjusted for demographic variables mainly via the microdata, our analysis includes a population-level variable, **dependency ratio**, measuring the ratio of the population aged 0 to 14 or 65 and older to the working-age population aged 15 to 64. This variable captures how much pressure the population exerts on public services.

Our confounder set includes three measures for public spending: **Education spending** as a percentage of total government spending; **health spending** as a percentage of total government spending, and; **total government spending,** as percentage of GDP. Public policies can be measured with spending measures and institutional qualitative indicators of policies (e.g., family policies, or unemployment insurance indicators) (Beckfield 2018). In the absence of complete institutional indicators, we selected public spending measures to capture how much resource governments allocate to public services that will directly affect children's likelihood of deprivation.

We include central political and institutional measures (Daoud 2015; Halleröd et al. 2013; Rothstein 2014). **Democracy**: Autocratic regimes can solicit the IMF's advice with less political cost compared to societies that are more democratic. **Political terror**: a proxy measuring how likely social movements could mobilize to change public policies. **Government efficiency**: an indirect measure of how effective a government is in implementing public policies. **Corruption**: more corruption is likely to divert resources from critical public services. **Minimum age labor law**: a binary variable indicating whether this law is in place regulating child labor. During economic turmoil, poorer families are more likely to let their children work. **International war** and **Civil war**: the IMF avoid countries engaged in conflict.

However, at least one critical confounder is not directly observable: a government's political will to implement IMF programs. As these programs contain policies that are often difficult to implement, political will is an important driver of selection bias. A government can opt to select into an IMF program or policy, making the exposure a non-random exposure assignment. As discussed in the Appendix, we us an Heckman selection model to indirectly measure political will.

# 3    Results

## 3.1    Measuring IMF-education policy

Of the 138 countries present in the IMF conditionality dataset, 50 countries had at least one IMF-education policy listed in their EBS document between the period 1985-2014. Of the 67 countries in our sample, 39 had IMF programs. Of these 39, we find that 18 had at least one IMF-education policy. Counting over the entire period, Sierra Leone received most IMF-education policies: 23. These policies ranged from arguably small interventions—in terms of its potential effect on children's prospects to complete school—such as "complete and verify nationwide teacher's census" listed in its 1997 program, to major ones, "Identify and implement concrete measures to control the teacher's payroll budget" in its 2002 program.

Our search identified a total of 137 IMF-education policies—all listed in the Supplementary data. Although this is a small number compared to the over 50 000 conditions that the IMF has issued, even one IMF-education policy can have severe implications for children's chances for completing school their content. In Kenya 1989, an IMF-education policy stipulated "fiscal measures in the context of the 1989/90 budget, including user charges in the health, education, and other sectors." In Bulgaria 2006, a policy required the "Implementation of an employment cut of at least 5,500 positions in the education sector." In Tajikistan 2004, the IMF wrote "Reduce the number of employees in the education sector by 5 percent." Moreover, IMF-education policies affect not only children but also university students. In another of Kenya's IMF programs in 2001, a policy read, "Implementation of tuition and direct charges for university students as well as reduction in amount of the student loan." In Bolivia 2000, "Education: Develop a reform proposal for higher education in order to reduce the share of public resources for higher education." Consequently, even if IMF-education policies are few in numbers, when implemented, they are likely to have profound consequences to children and university students.

## 3.2    Evaluating the effect of IMF-education policies on children's chances of completing school

Based on the IMF-education policy data, we proceed to analyze the impact of IMF-education policies on children's educational prospects. The analysis will also evaluate the effect of IMF programs, revealing the extent to which it is educational policies or the set of all policies that affect children. Table 1 shows the raw covariate difference between children living in countries with an IMF program containing at least one IMF-education policy versus those that have no such programs. Similarly, Table 2 compares covariate balance by IMF program. For both exposures, covariate balance is fairly comparable across the exposed and non-exposed groups.

[Table 2 about here]

Figure 2 panel (a), shows the ATE estimates for IMF-education policies and IMF programs. While the estimate of IMF-education policy has an adverse effect on children's education deprivation, the effect is not significant at a 95%-confidence interval. The estimate suggests an adverse increase in deprivation by about half a percentage point. One statistical reason for which the effect is insignificant is that the model is underpowered: the data contains only 14 countries with at least on IMF-education policy. A substantive reason for the lack of effect is that IMF programs contain a slew of other policy conditions that dominate any potential effect of IMF-educational policies. Corroborating that reason, the model estimates an adverse IMF-program effect of 6 percentage points (with a standard error of 0.028), shown in panel (b). This effect means that in a country with no IMF program and which would implement such a program will on average see an increase in children failing to complete school by a proportion of six. In a population with 100 children and zero education deprivation to before an IMF program, 6 children will fail after implementations. Thus, keeping the statistical caveat that the IMF-education data is possibly underpowered, our results show that IMF programs (as a set of policies) carry more political power to affect children's chances of completing school.

Figure 3 shows the effect of IMF programs by children's age group.[2] Following the trends of the two ATEs, these disaggregated effects reveal that IMF programs induce an adverse effect on all groups. These effects range from 9 (age seven) to 4 (age eleven) percentage points: hovering around 5 points. For younger children (seven to eleven), our models cannot verify that their respective estimates are different from zero at a 95%-confidence interval. For older children (twelve to seventeen), the effects are significantly different from a zero estimate for children aged twelve to fourteen and sixteen; for the fifteen-years and seven-years old children, the effect is only significant at a 90%-confidence interval. These differences in significance are likely to the power of the data—the age-by-group samples are of different size. As presented in the Discussion, one likely explanation to this finding—that IMF programs are likely affecting the older group more than the younger one—is because the education system for older children required more resources than for younger children. In times of austerity, governments likely restrain education spending.

ATE estimates reveal informative child-population-wide trends, yet they mask how the effect varies across groups of children. CATE is useful for unpacking these variations. Figure 4 shows CATE distribution of estimates for each child (across all ages), revealing a wide variation across the IMF-program ATE. For some child subgroups, IMF programs are likely not affecting them (those around the red line) at all or even improving their situation (those below the redline). To better understand why some children, react differently to IMF programs, we analyze subgroups of children based on the five quantiles (i.e., quintiles) of the CATE distribution.

A useful property of the GRF model is that it suggests which variables in the covariate set moderate most of the CATE distribution. Figure 5 shows the top-ranking moderators, their average value, and their variation by quintiles. The analysis focuses on the model for seventeen-years-old children, as this group of children are the ones that policy changes are likely affect their living conditions, beyond completing school. They are the ones that are on the verge of leaving childhood and moving into adulthood—with all the complexities that movement entails, from finding a job, shaping a new home, and forming a family. Completing this movement with or without an education makes a large difference.

---

[2] The IMF-education policy effect follows a similar but statistically weaker estimate and so the analysis show IMF program effects.

The seven top-ranking moderators jointly capture about 80 percent of the CATE variation. For this top group, Figure 5 start with listing the highest-ranking moderators and end with the lowest. The GRF model suggest that their moderating differences are small, they range in about 7 to 10 in importance value. There are at least three striking observations. First, children with the highest impact—those in fifth quantile (Q5)—live in countries with high dependency ratio. A value of 82 for this ratio means that 82 percent of the population are between 0-14 or 65 and above. As our sample represents low- and middle-income countries and these countries tend to have a young population, it means that there are more children in these countries relative to adults. The more children, the higher the likelihood that children are affected by government cuts. While this finding is logical, the following two are not—at first sight.

Family wealth and government's education spending seems to have a reversed moderating effect on the relationship between the exposure and outcome. Second, family wealth variable ranges from one (poorest families) to five (wealthiest families). Families with a material-wealth value of 2.12 indicates that households with less resources are less affected by IMF programs (first quantile), and families of the middle-class with a value of 3.05 are most affected by IMF programs as they are in the fifth quantile of the CATE distribution. Third, governments that spend more on their education system prior to an IMF program, are also encountered with more children failing at school due to IMF programs.

By revisiting the content of IMF-education policies, we form a logical explanation to these counterintuitive findings. While our model is unable to quantitatively verify an effect of IMF-education policies, our qualitative analysis provides evidence that IMF-education policies matter for children's chances of completing school.

## 3.3 Qualitative analysis of IMF-policy documents: the link between IMF programs and children's educational outcomes.

The counterintuitive effect heterogeneity shown in Figure 5 has a partly causal and a partly associative explanation. The associative explanation emphasizes that education spending only appears to causally moderate child-education deprivation during IMF programs, when this moderation is merely a noncausal correlation. Education spending likely coincides with one or several unobserved factors that also produce more poor deprivation during IMF programs. One such unobserved factor is parents' unemployment status—a variable we excluded from our model because it would block the IMF effect as it is a post-exposure variable. Many governments worldwide endorse the importance of countercyclical policies (Stiglitz 2009). Consequently, in anticipating the onset of an economic crisis, or during such turmoil but before the involvement of the IMF, governments are likely to have increased their education spending to cushion increasing rates of unemployment.

However, when the IMF becomes involved, high education spending merely appears correlated with education deprivation, even though unemployment causally explains both the increased number of children deprived and more education spending. Therefore, in the presence of IMF programs, unemployment is likely a confounder in the relationship between high education spending and children's education outcomes. In estimating the total IMF effect, our design controls for confounding in the causal relationship between IMF programs and education deprivation, not for confounding in the causal relationship between education spending and deprivation (VanderWeele 2015). To unravel this mediational effect, we would need a different research design—a key task for future research.

When we complement our statistical results with our qualitatively produced archival results, we find that the associative explanation cannot fully account for the links among IMF programs, education spending, and education deprivation. To shed light on these links, based on our IMF-education-policy data, we further analyzed the content of IMF-policy documents for each country in our study.

An archetypical example of an adverse IMF program is Tajikistan in 2000. In this program, the IMF and the government agreed to "reduce the number of employees in the education sector by 5 percent," aiming at increasing "the public sector wage bill by 28 percent" (IMF 2003:13), including setting teachers' salaries "on a merit basis" (IMF 2003:14). However, this 5 percent reduction turns out to merely constitute the beginning of a larger plan to entirely reform the education system. Indeed, the government of Tajikistan committed to the following IMF policy adjustments:

> We will reduce the number of employees in the education sector by 30 percent over a period of three to five years. We will begin the process of downsizing by reducing the number of budgetary employees in the education sector by 5 percent as of July 2004 (structural benchmark). Further, the plan will address reform of the education budget system with a move to more school autonomy; introduction of a fee schedule; expansion of teaching assignments; curriculum reform; and private sector involvement. (IMF 2003:69)

Similarly, in Bolivia's 1999 program, IMF policies stipulated that the Bolivian government had to "develop a reform proposal for higher education in order to reduce the share of public resources for higher education" (IMF 1999:73). The reason for this reform was that the Bolivian government "had done little to smooth income distribution [as…] much of government social spending fails to reach the poorest groups of society. A disproportionate share of public spending on education goes to universities" (IMF 1999:12). In other words, the IMF and the Bolivian government agreed to directly target low-income families by reallocating a portion of higher education spending in favor of rural-development programs (IMF 1999:62). While such reforms may help the rural population, this reform adversely affects the middle class, because they attend higher-education at a higher rate than the poorer segments of society (Buchmann 1996; Daoud and Puaca 2011).

Nonetheless, how does cutting higher-education funding at $t-1$ that affects mainly young adults (age $\geq 18$), increase the probability of material deprivation for children between the ages 0 to 17 at $t$? One likely pathway is that these cuts affect young adults that are already attending, or planning to attend, university and that these adults also have children that indirectly are affected by these cuts. In Bolivia, the IMF suggested "raising the tuition fees charged by universities." (IMF 1999:13). Raising fees put a financial burden on the families of these young adults as they must start paying more for acquiring a university education than they did before the reform (Alexander 2001; Buchmann 1996). Consequently, such increased university fees affect adversely young adults and their children's risk of failing in school.

Another likely pathway is via a sibling effect. Older siblings (above the age of 18) who would have attended higher universities now choose to search for a job but find themselves unemployed instead. As the youth unemployment rate in low and middle-income countries is high (Banerjee and Duflo 2012), chances of finding a job is low (Buchmann 1996; Daoud and Puaca 2011). If the older sibling remains unemployed, then that indicates that the older sibling will likely not move out of his parents' home and has to keep sharing scarce resources with their younger sibling (Daoud 2007, 2010, 2018; Garfinkel, McLanahan, and Wimer 2016). This sharing may burden the younger sibling by diverting resources away from them and thus increasing the younger sibling's risk of education deprivation. These observations reinforce our first key finding that IMF reforms affect the middle class at least as much as low-income families but through different causal pathways.

Both Tajikistan and Bolivia spent about the sample average (third quantile) on education before selecting into IMF programs. The combination of spending on education, presence of IMF-education policies, and effect on child-education deprivation provide a clue to the likely causal relationship. In high- and medium-education spending countries, the IMF will likely identify a larger fiscal leeway for cutting this spending to balance fiscal deficit than for countries with low education spending. Subsequently, such cutting will increase deprivation. If our reasoning holds empirically, it would suggest that low-education-spending countries have fewer or no adverse IMF-education policies. Indeed, that is also what we find.

In some countries with low education spending before IMF program enrollment, such as Gabon and Chad, the IMF introduced education spending floors. In Chad, the IMF and government officials decided that they would "follow pro-poor expenditure more closely, indicative minimum targets are set for education and health current spending; a ceiling on the government's total wage bill is also set, with a view to ensuring the attainment of the program's fiscal objectives" (IMF 2001:19). These types of pro-poor policies, while paying attention to the fiscal ceiling, are common in low-income countries and often set in collaboration with the World Bank. Nonetheless, Kentikelenis et al. (2016) find that these spending targets have low priority in the slew of other high-priority IMF policies and thus tend to be inadequately implemented—if at all.

## 4    Discussion

Based on natural language processing, we identified the existence which IMF policies target a country's education system (Åkerström et al. 2019; Daoud, Reinsberg, et al. 2019). While IMF-education policies exist in several IMF programs, our statistical model finds no effect of these educational policies on children's probability of completing school. The lack of effect of IMF-education policies is likely driven by that these policies are dependent on the presence of an IMF program. As IMF-education policies cannot exist without the presence of an IMF program, this dependence implies that statistically any IMF impact on children is mixed with other IMF policies. As IMF programs are dominated by macroeconomic policy conditions, it is likely that these conditions drown out any potential direct effect from IMF-education polices. But this dependence is a consequence of how IMF programs are designed: IMF and government officials design a program to mainly address critical macroeconomic issues rather than issues in the education system. Focusing on macroeconomics will therefore affect the educational system via indirect channels, mainly reduced education spending. Nevertheless, as our archival analysis shows, albeit in small numbers, the IMF requires direct targeting the education system. This mix of direct and indirect targeting are likely the mechanisms through which children are affected. Consequently, as IMF-education policies are predominately bundled into a set of other policies that make up an IMF programs, children are affected by the presence or absence of austerity rather

Our analysis shows that IMF programs affect mainly children between the age of 12 to 17. Schooling for younger children, between the age of 7 and 11, is less resources heavy than for older children. These age groups that the model detects matches with the age groups countries use around the world, although with small differences. Maintained by the United Nations Educational, Scientific and Cultural Organization (UNESCO), the International Standard Classification of Education (ISCED) defines primary education as the education stage suitable for children from age six to twelve and secondary education for children in the age range thirteen to nineteen. These thresholds matches well with our our households data that covers the ages seven to seventeen. Compared to primary education, schooling for secondary education requires more teachers (often with higher salary), increasingly expensive pedagogical material, and a variety of other educational resources. This makes schooling for older children more dependent on sufficient education spending. Thus, a government must spend considerable amounts to maintain high educational standards. Despite that IMF-education policies directly target the educational system, this finding—that the older children rather than the younger are most effected by IMF programs—substantiates the argument that it is the mix of direct and indirect policies aggravates children's risk of completing secondary school.

# 6 Tables

*Table 1. Covariate balance stratified by IMF exposure status*

|  | IMF = 0 | IMF =1 |
|---|---|---|
| n | 985805 | 955929 |
| Child age (mean (sd)) | 8.27 (5.06) | 8.08 (5.07) |
| Child sex = Male (%) | 501117 (50.8) | 483457 (50.6) |
| Urban household = rural (%) | 613439 (62.2) | 603594 (63.1) |
| Family wealth (%) | | |
| 1 | 256662 (26.0) | 223967 (23.4) |
| 2 | 192713 (19.5) | 192630 (20.2) |
| 3 | 187835 (19.1) | 190872 (20.0) |
| 4 | 170723 (17.3) | 173794 (18.2) |
| 5 | 177872 (18.0) | 174666 (18.3) |
| Nr of children (mean (sd)) | 3.77 (2.44) | 4.04 (3.21) |
| Nr of adults (mean (sd)) | 2.91 (1.87) | 2.97 (2.42) |
| Head of household education (%) | | |
| noEducation | 314602 (31.9) | 273742 (28.6) |
| Primary | 360369 (36.6) | 330722 (34.6) |
| SecondaryPlus | 310834 (31.5) | 351465 (36.8) |
| Year of interview (mean (sd)) | 2003.49 (3.68) | 2003.35 (2.72) |
| Health spending (mean (sd)) | 8.32 (5.61) | 10.03 (3.51) |
| Democracy (mean (sd)) | 5.12 (3.23) | 5.89 (2.01) |
| Trade (mean (sd)) | 66.07 (39.99) | 56.73 (27.38) |
| Expenses balance (mean (sd)) | -2.29 (6.31) | -3.40 (2.79) |
| Economic develop (mean (sd)) | 6.51 (0.86) | 6.09 (0.85) |
| War (mean (sd)) | 0.13 (0.80) | 0.21 (0.90) |
| Dependency ratio (mean (sd)) | 76.07 (11.95) | 80.31 (12.89) |
| Negative growth (mean (sd)) | 6.99 (6.82) | 5.04 (2.58) |
| Inflation (mean (sd)) | 0.10 (0.29) | 0.05 (0.22) |
| Education spending (mean (sd)) | 12.80 (5.97) | 15.15 (5.06) |
| Economic globalization (mean (sd)) | 41.26 (13.59) | 39.88 (13.12) |
| Gov. effectiveness (mean (sd)) | -0.53 (0.60) | -0.66 (0.36) |
| Political terror (mean (sd)) | 3.64 (0.91) | 3.22 (0.97) |
| Corruption (mean (sd)) | -0.66 (0.47) | -0.80 (0.46) |
| Child labor law (mean (sd)) | 0.51 (0.50) | 0.62 (0.49) |
| Public spending (mean (sd)) | 24.35 (8.44) | 22.12 (8.22) |

*Table 2 Covariate balance stratified by IMF-education exposure status*

|  | IMF-education = 0 | IMF-education =1 |
|---|---|---|
| n | 939576 | 181460 |
| Education deprivation (mean (SD)) | 0.14 (0.35) | 0.16 (0.36) |
| Child age (mean (SD)) | 11.73 (3.11) | 11.68 (3.11) |
| Child sex = Male (%) | 475713 (50.6) | 91907 (50.6) |
| Urban household = rural (%) | 589826 (62.8) | 104569 (57.6) |
| Family wealth (%) | | |
|   1 | 226930 (24.2) | 42841 (23.6) |
|   2 | 182837 (19.5) | 35725 (19.7) |
|   3 | 182537 (19.4) | 35557 (19.6) |
|   4 | 169102 (18.0) | 32248 (17.8) |
|   5 | 178170 (19.0) | 35089 (19.3) |
| Nr of children (mean (SD)) | 3.95 (2.86) | 4.19 (2.70) |
| Nr of adults (mean (SD)) | 3.00 (2.21) | 2.88 (1.74) |
| Head education (%) | | |
|   noEducation | 276459 (29.4) | 62645 (34.5) |
|   Primary | 338164 (36.0) | 60226 (33.2) |
|   SecondaryPlus | 324953 (34.6) | 58589 (32.3) |
| Year of interview (mean (SD)) | 2003.66 (3.36) | 2002.83 (2.41) |
| Health spending (mean (SD)) | 8.78 (4.67) | 11.84 (3.80) |
| Democracy (mean (SD)) | 5.72 (2.70) | 5.40 (2.22) |
| Trade (mean (SD)) | 62.85 (35.37) | 52.52 (31.85) |
| Expenses balance (mean (SD)) | -3.42 (4.73) | -2.43 (1.92) |
| Economic develop (mean (SD)) | 6.32 (0.88) | 6.28 (0.93) |
| War (mean (SD)) | 0.19 (0.90) | 0.00 (0.00) |
| Dependency ratio (mean (SD)) | 76.64 (12.64) | 80.90 (12.02) |
| Negative growth (mean (SD)) | 5.21 (2.44) | 5.83 (2.13) |
| Inflation (mean (SD)) | 0.08 (0.27) | 0.08 (0.27) |
| Political will (mean (SD)) | -0.14 (0.50) | 0.15 (0.43) |
| Education spending (mean (SD)) | 14.48 (5.50) | 13.59 (3.38) |
| Economic globalization (mean (SD)) | 41.13 (12.95) | 41.53 (14.18) |
| Gov effectivness (mean (SD)) | -0.54 (0.48) | -0.59 (0.37) |
| Political terror (mean (SD)) | 3.39 (0.89) | 3.48 (1.12) |
| Corruption (mean (SD)) | -0.72 (0.48) | -0.65 (0.41) |
| Child labor law (mean (SD)) | 0.53 (0.50) | 0.66 (0.47) |
| Public spending (mean (SD)) | 23.54 (8.71) | 21.84 (6.60) |
| IMF program (mean (SD)) | 0.41 (0.49) | 0.93 (0.25) |
| Countries with IMF (mean (SD)) | 58.25 (8.27) | 62.45 (6.79) |
| UN vote G7 (mean (SD)) | 0.62 (0.05) | 0.62 (0.05) |
| IMFeduc (mean (SD)) | 0.00 (0.00) | 1.00 (0.00) |

# 7 Figures

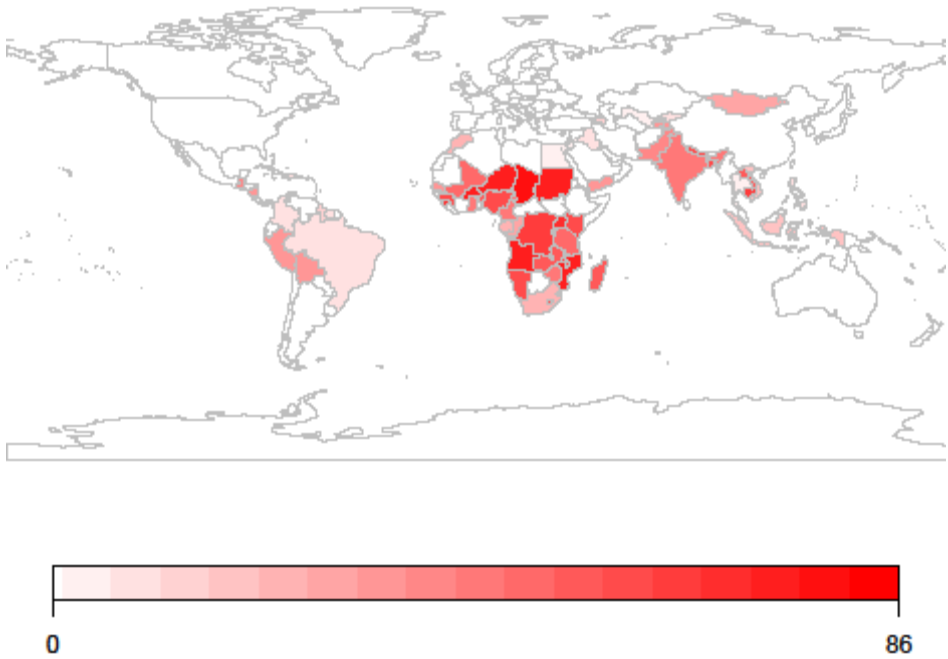

*Figure 1. Global distribution of child-education deprivation.* Notes: we calculated country averages from the DHS and MICS micro data for each of the 67 nations included in the sample. White colored countries are not included in the sample.

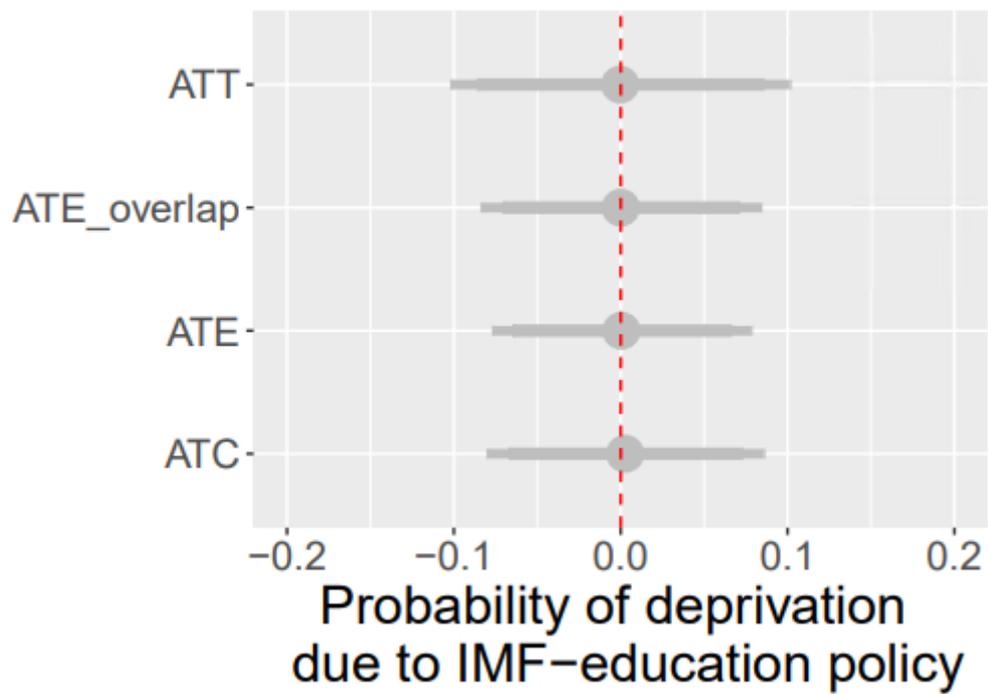

Panel a

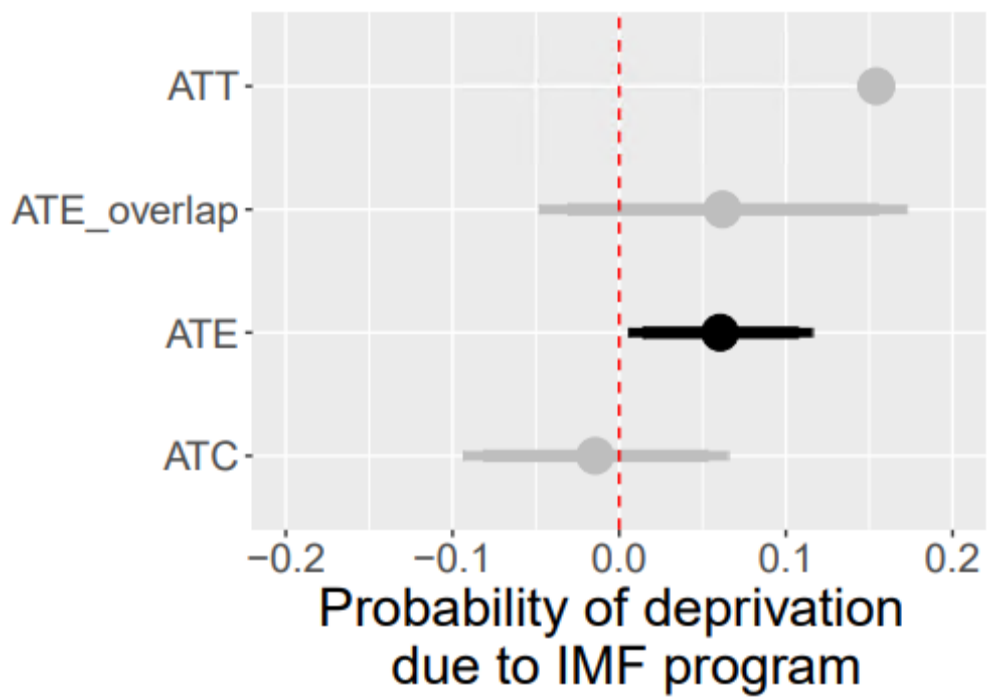

Panel b

*Figure 2: ATE for IMF-education policy*

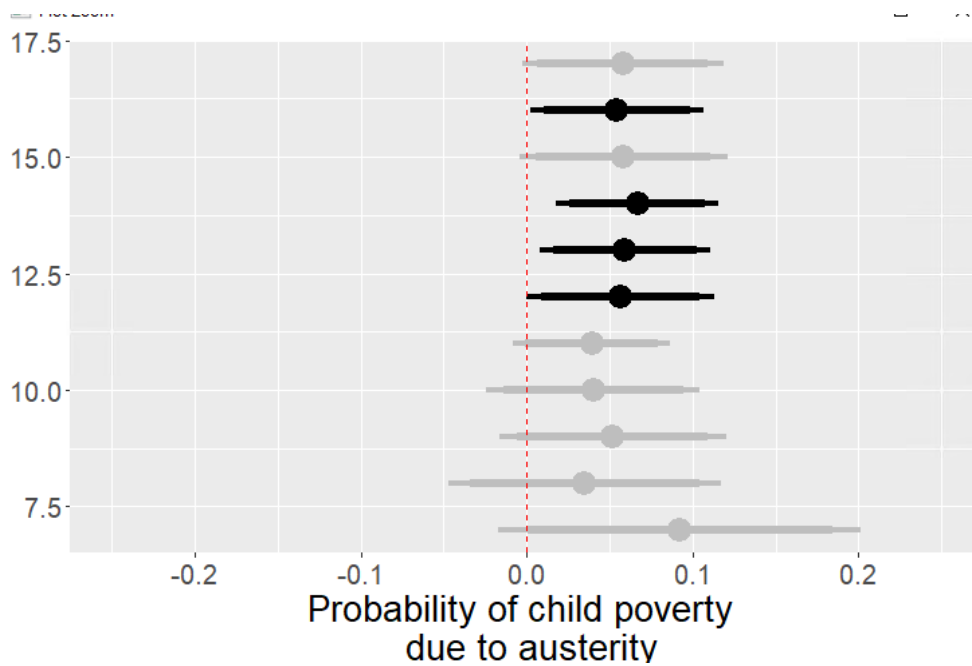

*Figure 3: ATE for IMF program by child age*

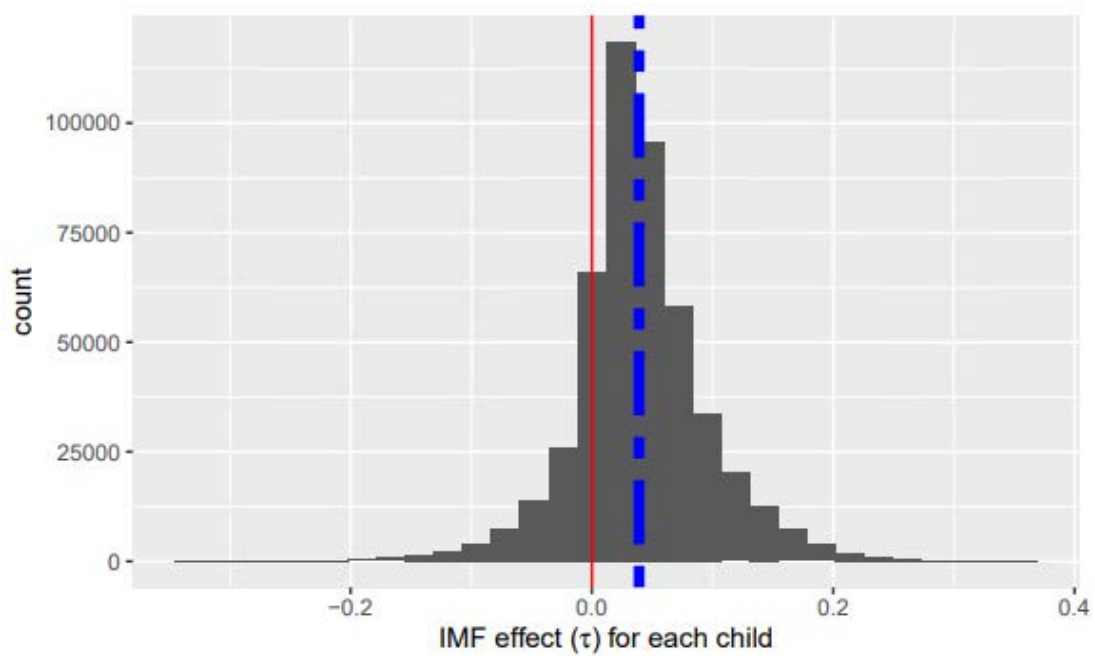

*Figure 4: CATE distribution for all children (age 7-17)*

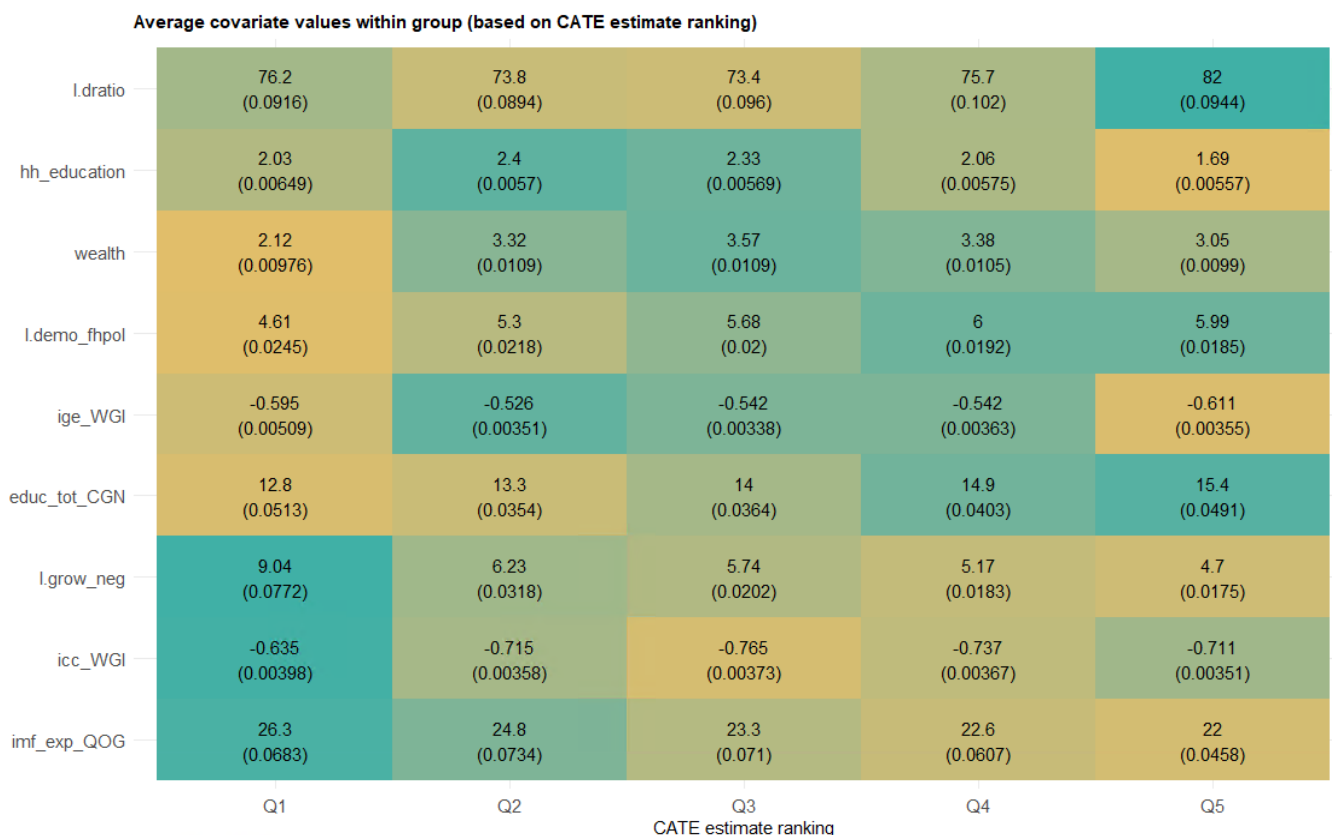

*Figure 5: Group characteristics by quantile ATE for IMF program by child age*

# 8 Appendix

## 8.1 Appendix A: Measures

| Variable | Definition | Source |
|---|---|---|
| *Country-level variables* | | |
| **Health spending** | Measured as a share of GDP, and as a share of total government expenditures. | IMF, 2011, and World Development Indicators |
| **IMF program** | Dummy variable produced by the IMF: 'the starting year of an IMF-supported program [is defined] as the year in which the program was approved, provided this occurred in the first half of the year. If the approval date was in the second half of the year, the starting year is the following year. The end year is the year in which the program | IMF, 2011. |

expired.'

| | | |
|---|---|---|
| **GDP per capita** | Gross domestic product per capita (constant 2000 USD) – logged (to correct for the skewed distribution). | WDI, Sep. 2012. |
| **Government balance** | General government balance (share of GDP). Calculated by subtracting general government expenditure from general government revenue. | Authors' calculation using IMF-WEO data. |
| **High inflation** | Dummy variable: = 1 if year-to-year change in inflation over 20%, 0 otherwise. | Authors' calculation using IMF-WEO data. |
| **Dependency ratio** | Combined shares of populations aged 0-14 and 65 and above. | Authors' calculations using WDI data. |
| **Trade** | Trade is the sum of exports and imports of goods and services measured as a share of gross domestic product. | WDI, Sep. 2012 |
| **Democracy** | Democracy, Range: 0-10 (Freedom House/Imputed Polity). Average of Freedom House and Polity (transformed to a scale 0-10). Hadenius & Teorell (2005) show that this average index performs better both in terms of validity and reliability than its constituent parts. | Quality of Governance Database, 2011. |
| **Negative growth** | Dummy variable: = 1 if negative growth in a given year, 0 otherwise. | Authors' calculation using IMF-WEO data. |
| **ODA** | Net ODA received (% of GNI). | WDI, Sep. 2012 |
| **Low income country** | Dummy variable. Country is eligible for concessional lending from the IMF | IMF, 2011 |
| **Sub-Saharan Africa** | Dummy variable. Refers to countries located south of the Sahara Desert. | World Development Indicators |
| **Civil war** | Magnitude score of episode(s) of civil warfare | Center for Systemic Peace |
| **Education expenditure (% of GDP)** | Public education spending as a percentage of GDP | Clements, Gupta & Nozaki 2011, What happens to social spending in IMF-Supported programs? |
| **Education expenditure (% of gov. spending)** | Public education spending as a percentage of total government spending | Clements, Gupta & Nozaki 2011, What happens to social spending in IMF-Supported programs? |
| **Economic globalization** | The KOF Globalisation Index measures the economic, social and political dimensions of globalisation. | Economic globalization, Dreher 2006, KOF Index of Globalization (Version: March 2016) |
| **Government effectiveness** | "Combines into a single measure of the quality of public service provision, the quality of the bureaucracy, the competence of civil servants, the independence of the civil service from political pressures, and the credibility of the government's commitment to policies. The index focus on the "inputs" required for the government to be able to produce and implement good policies and deliver public goods". (Teorell et al. 2018) | Interpolated, World Bank, Worldwide Governance Indicators |
| **Corruption** | "measures perceptions of corruption, defined as the exercise of public power for private gain. It measures aspects of corruption ranging from the frequency of "additional payments to get things done", to the effects of corruption on the business environment, to measuring "grand corruption" in the political arena or in the tendency of elite forms to engage in "state capture". (Teorell et al. 2018) | Interpolated, World Bank, Worldwide Governance Indicators. |

| **Political terror** | Measures political terror on a scale 1 to 5, where 5 is the most severe form of terror. This level is defined as, "Terror has expanded to the whole population. The leaders of these societies place no limits on the means or thoroughness with which they pursue personal or ideological goals" (Teorell et al. 2018) | Political terror scale (U.S. State Department), Gibney, Cornett & Wood 2013, Political Terror Scale.. |
|---|---|---|
| **Minimum age labor law** | Dummy variable: = 1 a law in place regulating minimum required working age , 0 otherwise | Minimum Age Convention, 1973 (No. 138), International Labour Organization (ILO), Information System on International Labour Standards (NORMLEX) (Retrieved 2014: http://www.ilo.org/normlex |
| **Government spending** | Government expenditure (Percent of GDP) | IMF in Quality of Governance Database, 2011 |
| _Family-level variables_ | | |
| **Nr children** | Number of individuals under the age of 18 | Demographic and Health Survey; Multiple Indicator Cluster Survey |
| **Nr of Adults** | Number of individuals over the age of 18 | Demographic and Health Survey; Multiple Indicator Cluster Survey |
| **Education** | Ordinal variable (no education, primary, and secondary+). Measures the head of household's level of education. | Demographic and Health Survey; Multiple Indicator Cluster Survey |
| **Wealth index** | Ordinal variable (Quintiles). The index is a composite measure of the household's material standard. It is calculated from selected assets such as ownership of | Demographic and Health Survey; Multiple Indicator Cluster Survey. (Rutstein 2008) |

| | television, mobile phones, bicycles. | |
| --- | --- | --- |
| **Urban rural** | Dummy variables. Captures the geographical location of households. | Demographic and Health Survey; Multiple Indicator Cluster Survey |
| *Child-level variables* | | |
| **Severe child health deprivation** | Dummy variable. Children under the age of 5 who had not been immunized against diseases or had a recent illness involving diarrhea and had not received any medical advice or treatment two weeks prior to the survey | Demographic and Health Survey; Multiple Indicator Cluster Survey. (Gordon et al. 2003:8) |
| **Sex of the child** | Dummy variable. | Demographic and Health Survey; Multiple Indicator Cluster Survey |
| **Age** | Age of the child in number of years. | Demographic and Health Survey; Multiple Indicator Cluster Survey |

## 8.2 Appendix B: Heckman Selection model

Governments select into IMF programs. This produces selection bias where countries that are rapidly getting poorer might be more likely to cooperate with the IMF. If so, increases in poverty would be correlated with IMF intervention, even if the intervention did not cause increases in poverty. While observable variables affecting both selection into an IMF program and child poverty are already included as controls in our model, we cannot directly control for unobservable factors such as 'political will', as outlined in our DAG.

Four approaches have been used in the IMF program evaluation literature to address this type of selection bias: matching methods; instrumental variable approaches; system GMM estimation; and Heckman selection models. For our purposes, Heckman's two-step method is the most suitable choice to address concerns of selection bias as it produces a proxy for unobserved factors that we can include into our set of controls. The Heckman model involves first modelling IMF participation, and second modelling the outcome of interest using the inverse Mills ratio from the first step. Accordingly, in the first step, we estimate a probit model to predict the likelihood of IMF participation:

$$\Phi(Z_{k,t}\gamma) = Probit(imf.prog.cgn_{k,t}) = \gamma_0 + \gamma_1 imf.prog.cgn_{k,t-1} + \gamma_2 gdp.growth_{k,t-1} +$$
$$\gamma_3 cab.gdp_{k,t-1} + \gamma_4 demo.fhpol_{k,t-1} + \gamma_5 lngdppc_{k,t-1} + \gamma_6 civilwar_{k,t-1} +$$
$$\gamma_7 int.war_{k,t-1} + \gamma_9 UNvoteG7_{k,t-1} + \gamma_{10} CountriesWithIMF_{k,t-1}$$

As a point of reference, we rely on a version of the specification suggested by the Independent Evaluation Office of the IMF (IEO 2003): one that retain the best data coverage but which still gives analogous results. The outcome variable, $imf.prog.cgn_{k,t}$, measures if country $k$ had an IMF program at year $t$. Our choice of which central mechanism affect selection into programs, builds on Moser and Strum's suggestions (Moser and Sturm 2011):

- Previous IMF participation (imf.prog.cgn, t-1): a country's past involvement with the IMF tend to positively determine future program participation. The nearer historically, the more likely participation is. We use whether the country had a program last year.
- GDP growth (gdp.growth): Countries with lower growth are more likely to become economically constrained, and ask for IMF credit.
- Current account balance (cab.gdp as share of GDP): One of the key objective of the IMF is to support countries to overcome balance of payment issues deriving from trade. The higher the imbalance, the more likely the country is to ask for IMF help.
- Democracy (demo.fhpol): Autocratic regimes can with less political cost invite the IMF, compared to more democratic countries.
- Log GDP per capita (lngdppc): low income countries tend to seek concessional IMF assistance, whereas middle income countries with short term economic disturbances (e.g. currency crisis) tend to ask for non-concessional loans (e.g. Brazil, Argentina).
- Civil war (civilwar): Even if countries with a high degree of domestic civil conflict might need more economic help, the IMF might avoid involvement during violent periods. Also, the political cost to call for IMF assistance might be high.
- International war (int.war): Countries involved in armed conflicts between sovereign nations deters the IMF.
- UN votes with G7 (UN vote G7): this variable captures how often countries vote in line with G7. This shows political proximity with the key nations driving the IMF.
- Countries on IMF programs (CountriesWithIMF): In any given year, the more countries that have IMF funding, the less likely the IMF is to issue new programs as its funds are limited.

The total number of countries on IMF programmes and UN voting patterns with G7 act as our exclusion restrictions: these variable explain significantly the country's participation decision in IMF programs but are not correlated with the dependent variable of the outcome equation, in our case child poverty. Voting pattern has stronger relevance as it is significantly correlated in all alternative selection specifications.

We choose not to include government balance (lagged one year) as it reduced many observations due to missing data. We would lose 6 countries reducing our data size by 10%. We calculate the inverse Mills ratio and include it in the outcome equation to control for the remaining unobserved variation (Heckman 1979). The equation below defines the inverse Mills ratio, $\lambda$, which isolates unobserved factors determining IMF participation:

$$\lambda_{k,t} = \begin{cases} \phi(Z_{k,t}\hat{\gamma})/\Phi(Z_{k,t}\hat{\gamma}), & \text{if } T_{k,t} = 1. \\ -\phi(Z_{k,t}\hat{\gamma})/(1 - \Phi(Z_{k,t}\hat{\gamma})), & \text{otherwise.} \end{cases}$$

The Mills ratio is calculated for each observation: country $k$ at time point $t$, and depending on their treatment status $T_j$ (present or absent IMF program). The function $\phi$ denotes the standard normal density function, and $\Phi$ the standard normal cumulative distribution function; $Z_{k,t}$ represents the covariates and $\hat{\gamma}$ are the vector of estimated parameter from the first equation. The inverse Mills ratio, $\lambda$, is then used as a covariate, in the outcome equation (in our case, the multilevel models with child poverty as outcomes) controlling for self-selection. In a linear model, its coefficient is interpreted as follows: if significantly negative, then unobserved variables that make IMF participation more likely are associated with lower government health expenditure; if significantly positive, then unobserved variables that make IMF participation more likely are associated with higher government health expenditure; if non-significant, then there is no association.

*Alternative selection specifications*

|  | *Dependent variable* IMF program (t) | | |
| --- | --- | --- | --- |
|  | M1 | M2 | M3 |
| IMF program (t-1) | 1.910*** | 1.959*** |  |
|  | (0.064) | (0.094) |  |
| GDP growth (t-1) | -0.018*** | -0.042*** | -0.026*** |
|  | (0.006) | (0.010) | (0.009) |
| Current account balance (t-1) | -0.007* | -0.008 | -0.008* |
|  | (0.004) | (0.005) | (0.005) |
| Democracy (t-1) | 0.027** | 0.033* | 0.046*** |
|  | (0.013) | (0.019) | (0.016) |
| Log GDP per capita (t-1) | -0.256*** | -0.253*** | -0.336*** |
|  | (0.034) | (0.053) | (0.054) |
| Log aid per capita (t-1) |  |  | 0.004 |
|  |  |  | (0.004) |
| Civil war (t-1) | -0.026 | 0.042 | 0.030 |
|  | (0.035) | (0.069) | (0.054) |
| International war (t-1) | 0.042 | -0.213 |  |
|  | (0.072) | (0.156) |  |
| Dependency ratio (t-1) |  |  | -0.002 |
|  |  |  | (0.004) |
| Countries on IMF programs | 0.011*** | 0.006 | 0.001 |
|  | (0.004) | (0.005) | (0.005) |
| UN voting pattern with G7 | 0.886** | 1.041* | 3.303*** |
|                    | (0.419)   | (0.546)  | (0.686)  |
| Constant           | -0.570*   | -0.418   | -0.130   |
|                    | (0.309)   | (0.478)  | (0.741)  |
| Observations       | 2,482     | 1,264    | 1,066    |
| Log Likelihood     | -1,000.302| -471.559 | -631.869 |
| Akaike Inf. Crit.  | 2,020.605 | 963.118  | 1,283.739|
| *Note: probit models* | *p<0.1, **p<0.05, ***p<0.01 | | |